\documentclass[twocolumn,prb,floatfix,showpacs]{revtex4}

\usepackage{epsfig}
\usepackage{epsf}

\newcommand{\beq}{\begin{equation}}
\newcommand{\eeq}{\end{equation}}
\newcommand{\beqq}{\begin{eqnarray*}}
\newcommand{\eeqq}{\end{eqnarray*}}
\newcommand{\bra}{\langle}
\newcommand{\ket}{\rangle}

\begin{document}

\title{ Effect of Semicore Orbitals on the Electronic Band Gaps of 
Si, Ge, and GaAs within the GW Approximation}

\author{ Murilo L. Tiago, Sohrab Ismail-Beigi, and
Steven G. Louie}

\address{Department of Physics, University of
California at Berkeley, Berkeley, California 94720-7300\\
and Materials Science Division, Lawrence Berkeley National Laboratory,
Berkeley, California, 94720}

\date{\today}

\pacs{71.15.-m, 71.15.Ap, 71.15.Dx, 71.10.-w}

\begin{abstract}
We study the effect of semicore states on the
self-energy corrections and electronic energy gaps of silicon,
germanium and GaAs. Self-energy effects are computed within the GW
approach, and electronic states are expanded in a plane-wave
basis. For these materials, we generate {\it ab initio}
pseudopotentials treating as valence states the outermost
two shells of atomic orbitals,
rather than only the outermost valence shell as in
traditional pseudopotential calculations. The resulting 
direct and indirect energy gaps are compared with
experimental measurements and with previous calculations
based on pseudopotential and ``all-electron'' approaches.
Our results show that, contrary to recent claims, self-energy effects due 
to semicore states on the band gaps can be well
accounted for in the standard valence-only
pseudopotential formalism.
\end{abstract}

\maketitle


\section{Introduction}

Since the early applications of the GW method to real materials (see
Refs. 1-3 and References therein), the pseudopotential plane-wave
approach has been the method of choice due to its 
accuracy and technical simplicity. Recent advances in LAPW and LMTO
methodologies have allowed the implementation of
``all-electron'' applications of the GW
method.\cite{blochl,hamada,kotani,ku,lebegue} One common feature of such
calculations, using standard level of approximation for the
self-energy, is an underestimation of the electronic energy
gap compared to experimental measurements, whereas
pseudopotential-based calculations show very good agreement with
experiment.\cite{h&l,wilkins,szl} To explain this inconsistency,
it was proposed that the pseudopotential approach does not correctly
describe the effect of core orbitals in the self-energy corrections
to the energy gaps,
resulting in overestimated corrections.\cite{kotani,ku,lebegue} 

It is thus desirable to elucidate the effect of core orbitals
in the quasiparticle band structure, and the preferred procedure
is to perform a well converged all-electron calculation and compare its results
with similarly converged pseudopotential-based
calculations. Obviously, numerical precision should not be neglected.
In this work, we explicitly include semicore orbitals
in the pseudopotential plane-wave approach and calculate the quasiparticle
energy gap for three semiconductors of technical importance: Si,
Ge and GaAs. The underlying description  of the ground-state electronic
structure is based on Density Functional Theory in the Local Density Approximation
(DFT/LDA).\cite{ihm,p-z} 
Throughout this work, we are careful to converge all results systematically,
and the final results are compared to previous
pseudopotential results and to recent all-electron calculations. The
paper is organized as follows: we outline the
theoretical method in section II. Results 
are presented in section III and discussed in section IV.

\begin{figure}
\centering\epsfig{figure=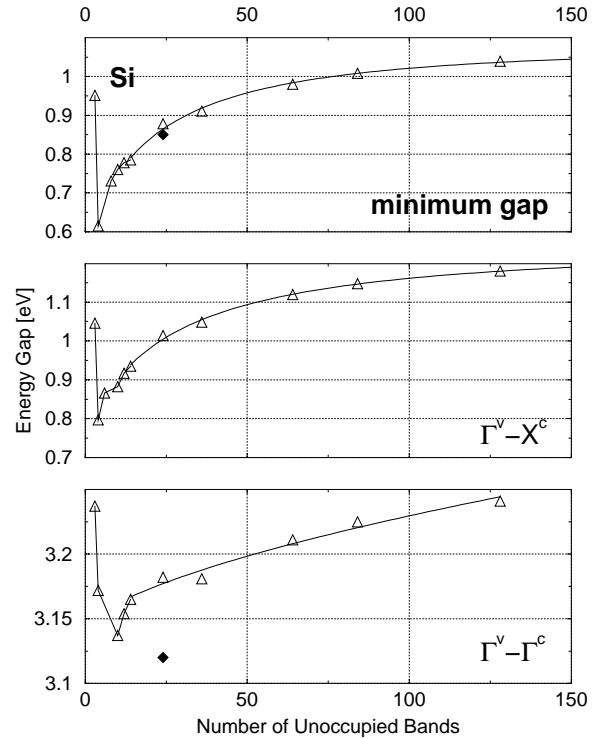,width=8cm,clip=}
\caption{Convergence of the minimum gap in Si as
  function of the number of unoccupied bands included in the
  calculation of the Green's function, Eq. (\ref{green}) (open
  triangles). The solid line is a guide to the eye. Results obtained by
  Ku and Eguiluz \cite{ku} are shown as black diamonds.}
\label{f.si}
\end{figure}


\section{Theoretical Method}

The pseudopotential formalism has two advantages that make it
convenient for practical {\it ab initio} calculations. First,
degrees of freedom due to core electrons are removed from the
system, resulting in a description that contains only valence electrons. 
Second, valence electronic pseudowavefunctions are
smooth in the vicinity of atomic sites as well as in the interstitial
region. Such valence wavefunctions can be expanded to convergence easily in a
relatively small basis set such as a plane-wave basis. On
the other hand, one possible problem with this formalism is that using 
pseudopotentials, instead of the true electron-ion
potential, may not fully describe all effects produced by
interactions between the valence and core electrons.
In order to address this issue in GW calculations, we include in the
present study electrons from the outermost
valence shell as well as those from the second outermost (``semicore'')
atomic shell as active ``valence'' electrons in the pseudopotential 
formalism. Only interactions
between these ``valence'' electrons and electrons
from the deeper core shells are described by pseudopotentials. Taking 
Si as an example, its ``core'' now contains only $1 s$
electrons. Electrons from the $2s$,$2p$,$3s$,$3p$, and $3d$ shells
are all treated on equal footing in the subsequent
calculations.

Apart from the atomic configuration, we follow the standard prescription 
for generating {\it ab initio} pseudopotentials. \cite{tm2,kerker} Since the
deeper core electrons have extremely large binding energy
({\it e.g.}, 130 Ry for the $1s$ electrons in atomic Si),
their interaction with valence electrons is expected to be
much weaker than the already small interaction between the outermost
valence and the semicore electrons.\cite{rohlfing}

The Kohn-Sham DFT formalism within the LDA is used to solve for the 
ground state electronic
structure and to provide a starting point for the calculation of the
electron self-energy. We follow closely the GW method as developed by
Hybertsen and Louie.\cite{h&l} In this method,
the self-energy is given by the standard GW approximation,

\begin{equation}
\Sigma ({\bf r}, {\bf r}^\prime;E) = i \int { {\rm d} E^\prime \over 2\pi} 
e^{-i E^\prime 0^+}
G_0 ({\bf r}, {\bf r}^\prime;E-E^\prime) 
W_0 ({\bf r}, {\bf r}^\prime;E^\prime) ,
\label{sigma}
\end{equation}
where $G_0$ is the one-electron Green's function, calculated from LDA energy
eigenvalues and eigenstates ($\delta = 0^+$ for occupied states,
$\delta = 0^-$ for unoccupied states),

\begin{equation}
G_0 ({\bf r}, {\bf r}^\prime;E) = \sum_{n{\bf k}}  {
 \varphi_{n{\bf k}} ({\bf r}) \varphi^\star_{n{\bf k}} ({\bf r}^\prime)
 \over E - \varepsilon_{n{\bf k}}^{\small DFT} -i \delta } \;\; .
\label{green}
\end{equation}

The screened Coulomb interaction $W_0$ is calculated within the Random
Phase Approximation (RPA) as $W_0 = \left[ 1 - V P_0 \right]^{-1} V$,
with $V$ being the bare Coulomb interaction and the polarizability 
$P_0 = -i G_0
G_0$.\cite{hedin} This is the commonly employed level of approximation
for GW calculations ({\it i.e.} neglecting self-consistency and vertex 
corrections) \cite{wilkins,aryasetiawan_rev}, 
and we specifically compare results from both pseudopotential
and all-electron calculations at this particular level of
approximation to help untangle the effects of core states.

\begin{table}
\caption{Atomic parameters used to generate semicore
  pseudopotentials. Pseudowavefunctions were defined from the orbital
  with lowest principal quantum number at each angular momentum channel.
  Cutoff radii are given in units of Bohr radius.}

\begin{tabular}{lccc}
\hline\\
  & Reference Configuration & $r_{cut}$
 & local channel\\
  &  & $s \hspace{0.8cm} p \hspace{0.8cm} d$ 
 & \\
\hline
Si & $ 2s^2 2p^6 3s^2 3p^{1.95} 3d^{0.05}$ & 0.40 \hspace{0.2cm} 0.35
\hspace{0.2cm} 0.40 & $d$ \\
Ge & $ 3s^2 3p^6 3d^{10} 
4s^2 4p^{1.5} 4d^{0.1}$ & 0.50 \hspace{0.2cm}
0.50 \hspace{0.2cm} 0.50 & $s$ \\
Ga & $ 3s^2 3p^6 3d^{10} 
4s^2 4p^1 4d^0 $ & 0.50 \hspace{0.2cm} 0.50
\hspace{0.2cm} 0.50 & $s$ \\
As & $ 3s^23p^6 3d^{10} 
4s^2 4p^2 4d^0 $ & 0.50 \hspace{0.2cm} 0.50
\hspace{0.2cm} 0.45 & $p$ \\ \hline
\end{tabular}
\label{t.atomic}
\end{table}

The convolution integral in Eq. (\ref{sigma}) is performed using the
generalized plasmon-pole (GPP) model,\cite{h&l} which enables one to
distinguish two contributions to the self-energy: a screened
exchange part ($\Sigma_{sx}$) arising from the poles
of $G_0$, and a dynamical Coulomb
interaction between an electron and the hole-like charge distribution
around it ($\Sigma_{ch}$) stemming from the poles of $W_0$. 
The former tends to increase the 
quasiparticle energy, after the bare exchange is excluded. The latter
tends to decrease the quasiparticle energy. In
particular, the dynamical contribution $\Sigma_{ch}$ is highly sensitive to
the number of bands $n$ included in the calculation of the Green's
function in Eq. (\ref{green}). The final quasiparticle energy of a state
$\varphi_{n{\bf k}}$ is given by ($V_{xc}$ is the LDA
exchange-correlation potential)

\begin{equation}
E_{n{\bf k}}^{qp} = \varepsilon_{n{\bf k}}^{\small DFT} + \bra \varphi_{n{\bf k}} |
 \: \Sigma ({\bf r}, {\bf r}^\prime;E_{n{\bf k}}^{qp}) \:
- V_{xc}({\bf r}) | \varphi_{n{\bf k}} \ket.
\label{eqp}
\end{equation}


\section{Results}

A semicore, non-relativistic pseudopotential was generated for
Si, using the
Troullier-Martins scheme.\cite{tm2} For Ga, Ge and
As, we constructed semi-relativistic pseudopotentials using the Kerker 
scheme.\cite{kerker} These choices resulted in stable, transferable 
pseudopotentials in the Kleinman-Bylander form,
without ghost states. \cite{k-b} 
A summary of atomic parameters is presented in
Table \ref{t.atomic}. A good expansion of electronic wavefunctions in
a plane-wave basis was obtained using a cutoff energy of 
700 Ry (600 Ry for Si), and the first Brillouin zone was sampled using 
a 4x4x4 Monkhorst-Pack grid. \cite{m-pack} These
numerical parameters ensure convergence in LDA energy eigenvalues to
0.01 eV or better. The Ceperley-Alder exchange-correlation potential
is used.\cite{p-z} 
For the lattice parameter, we used the experimental values:
5.43 $\AA$, 5.65 $\AA$, and 5.66 $\AA$ for Si,
Ge, and GaAs, respectively. The polarizability was 
expanded in a plane-wave
basis with an energy cutoff of 45 Ry (50 Ry for Si) and numerically
inverted for the calculation of 
the screened Coulomb interaction $W_0$. Numerical
precision in the calculation of the self-energy is 0.05 eV or better.

\begin{table}
\caption{Bands Gaps of Si. All quantities in eV.}

\begin{tabular}{lccc}
\hline\\
  & $E_{gap}$ & $\Gamma^v - \Gamma^c$
 & $\Gamma^v - X^c$\\
\hline
\multicolumn{4}{l}{LDA present work} \\
& 0.46 & 2.52 & 0.60\\
\hline 
\multicolumn{4}{l}{GW valence pseudopotential + CPP} \\
\phantom{XX} Shirley {\it et al.}$^a$ & 1.13 & 3.28 & 1.31 \\
\hline
\multicolumn{4}{l}{GW present work} \\
 & 1.04 & 3.24 & 1.18\\
\hline 
\multicolumn{4}{l}{GW all-electron} \\
\phantom{XX} Hamada {\it et al.}$^b$ & 1.01 & 3.30 & 1.14 \\
\phantom{XX} Kotani and van Schilfgaarde$^c$
& 0.89 & 3.12 &  \\
\phantom{XX} Ku and Eguiluz$^d$& 0.85 & 3.12 &  \\
\hline
Experiment$^e$ & 1.17 & 3.35 & 1.3 \\
\hline
\end{tabular}

\begin{flushleft}
$^a$ Reference 9.

$^b$ Reference 5.

$^c$ Reference 6.

$^d$ non-self-consistent results from Reference 7.

$^e$ Reference 23.
\end{flushleft}

\label{t.si}
\end{table}

Table \ref{t.si} shows some of the energy gaps obtained in the present
approach for Si,
compared with previous valence-only pseudopotential and
all-electron calculations. Overall agreement between the present
results (which explicitly include the effect of the semicore states)
and experimental measurements is
at the level of 0.1 eV, and discrepancies between our results and previous
pseudopotential-based calculations of Ref. 9 are equally small. 
Recent all-electron calculations carried out at the same level
of the GW approximation, however, 
systematically underestimate the minimum gap and the direct
$\Gamma - \Gamma$ gaps.\cite{kotani,ku} 
We find that the convergence of the self-energy with respect to the 
number of unoccupied bands included in $G_0$ in Eq. (\ref{green}) is 
an important factor. In Figure \ref{f.si}, we show the
behavior of the calculated energy gap as function of the number of
unoccupied bands, $n_c$, included in $G_0$. Convergence is typically
very slow, and
well-converged results require $n_c \geq 120$. Other energy
transitions ($\Gamma - X$ and $\Gamma - \Gamma$) show similar behavior
and also approach the converged value from below. In contrast, the
results of Ref. 7 were obtained with only $n_c = 24$
and are closer to our results at approximately the same value of $n_c$
than to the converged results, as shown in Figure \ref{f.si}.
This fact points to a lack of numerical convergence in 
the evaluation of the self-energy corrections in Ref. 7. 

\begin{figure}
\centering\epsfig{figure=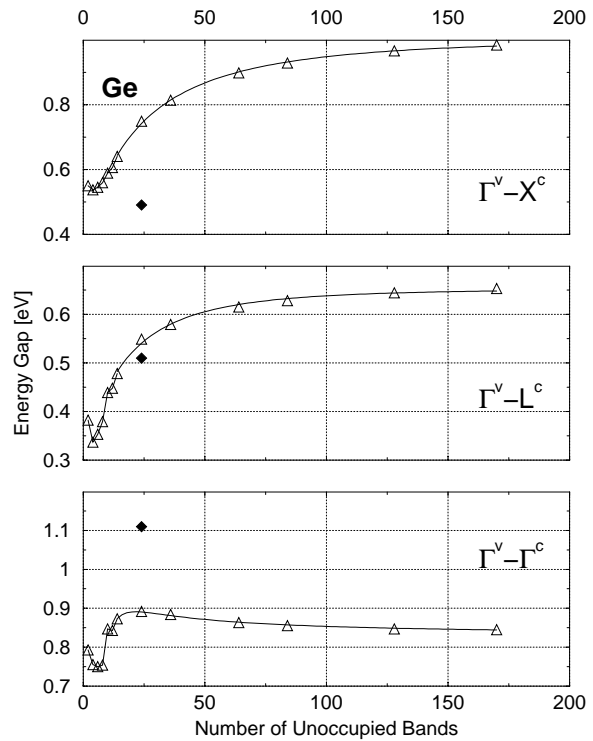,width=8cm,clip=}
\caption{Similar to Fig. \ref{f.si} for
  Ge. Numerical values for the
  gaps $\Gamma - \Gamma$ (upper panel), $\Gamma - L$ (middle panel),
  and $\Gamma - X$ (lower panel) are shown in open triangles. The full
  line is a guide to the eye. Results obtained by
  Ku and Eguiluz \cite{ku} are shown in black diamonds. }
\label{f.ge}
\end{figure}

\begin{table}
\caption{Bands Gaps of Ge. All quantities in eV.}
\begin{tabular}{lccc}
\hline\\
 & $\Gamma^v - L^c$ & $\Gamma^v - \Gamma^c$
 & $\Gamma^v - X^c$\\
\hline
\\
\multicolumn{4}{l}{LDA present work} \\
& -0.04 & -0.26 & 0.56 \\
\hline
\multicolumn{4}{l}{GW valence pseudopotential + CPP } \\
\phantom{XX} Shirley {\it et al.} $^a$ & 0.73 & 0.85 & 1.09 \\
\hline
\multicolumn{4}{l}{GW present work} \\
 & 0.65 & 0.85 & 0.98 \\
\hline 
GW all-electron &  &  &  \\
\phantom{XX} Kotani and van Schilfgaarde$^b$
& 0.47 & 0.79 \\
\phantom{XX} Ku and Eguiluz$^c$
& 0.51 & 1.11 & 0.49 \\
\hline
Experiment$^d$ & 0.74 & 0.90 & 1.3 \\
\hline
\end{tabular}
\vspace*{2ex}

\begin{flushleft}
$^a$ Reference 9.

$^b$ Reference 6 after inclusion of spin-orbit effects.

$^c$ non-self-consistent results from Reference 7.

$^d$ Reference 23.
\end{flushleft}

\label{t.ge}
\end{table}

The energy gaps obtained for Ge are presented in Table
\ref{t.ge}. Spin-orbit interactions are included as 
first-order perturbations.\cite{h&l-so} 
A common feature of LDA-based calculations is the
overlapping of the valence and conduction bands at the
$\Gamma$ point. The inclusion of $3 d$ electrons moves
the conduction bands further down resulting in a sizable negative direct 
gap.\cite{kotani,ku,szl,rohlfing,aryasetiawan_prb} This
feature is verified in our LDA calculation. As shown in Table \ref{t.ge}, 
self-energy corrections are
responsible for an opening of the gap and the correct positioning of
the minimum, indirect gap between points $\Gamma$ and $L$ in the Brillouin
zone. Our GW results compare well with experiment, although the
$\Gamma-X$ gap still shows a large discrepancy. On the other hand,
there are significant differences between the pseudopotential and
all-electron gaps at this level of the GW approximation, in particular
regarding the $\Gamma-X$ gap. As found in previous studies,
the self-energy corrections are needed 
to give the correct band topology. 

Figure \ref{f.ge} shows direct and indirect energy gaps in Ge as
function of the number of unoccupied bands included in the
calculation of $G_0$. Whereas the indirect gaps $\Gamma - L$ and $\Gamma -
X$ approach the converged value from below, the direct gap $\Gamma 
- \Gamma$ approaches it from above. This
particular convergence behavior arises from the fact that we are
plotting differences of quasi-particle energies: taken
individually, all quasi-particle energies $E_{n,{\bf k}}^{qp}$
converge monotonically from above, reflecting the attractive nature of
the Coulomb-hole $\Sigma_{ch}$ term.\cite{h&l,wilkins}
Additionaly, we note that in Ref. 7 the direct gap at $\Gamma$ is
strongly overestimated and the indirect gap $\Gamma-X$ is
underestimated by $\approx 0.8$ eV. As expected, the same pattern of 
overestimation/underestimation is evident in Figure \ref{f.ge} when 
we reduce the number of unoccupied bands from 170 to 24, which was the 
value used in Ref. 7.

\begin{figure}
\centering\epsfig{figure=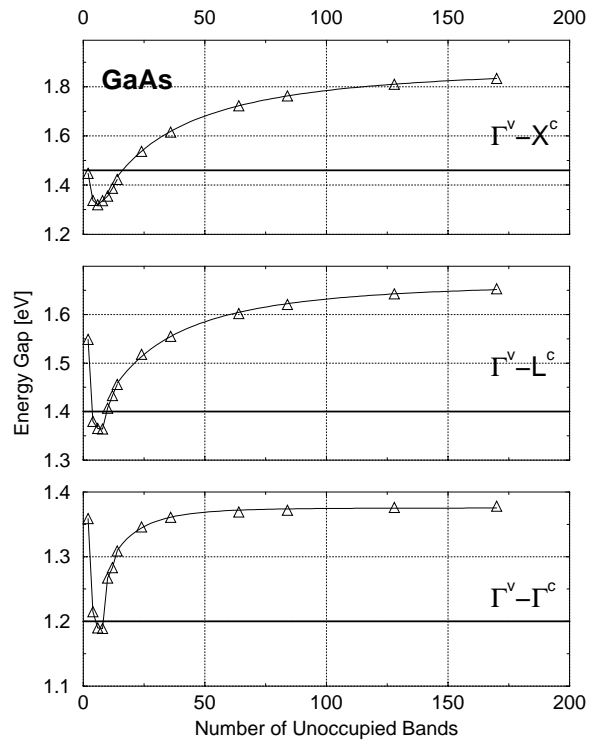,width=8cm,clip=}
\caption{Similar to Fig. \ref{f.ge} for GaAs. The 
 thick horizontal line represents results obtained by Kotani and
 van Schilfgaarde \cite{kotani} (the number of bands used was not
 reported in this reference). }
\label{f.gaas}
\end{figure}

GaAs shows behavior similar to Ge. Table
\ref{t.gaas} summarizes our results. Agreement with experimental data
is now within 0.15 eV, and all-electron calculations again
underestimate the energy gaps. Regarding the convergence
with $n_c$, Figure \ref{f.gaas} shows that the indirect gaps
converges slowly from below. On the other hand, the direct 
$\Gamma - \Gamma$ gap converges more quickly. 

\begin{table}
\caption{Bands Gaps of GaAs. All quantities in eV.}
\begin{tabular}{lccc}
\hline\\
 & $\Gamma^v - \Gamma^c$ & $\Gamma^v - L^c$
 & $\Gamma^v - X^c$\\
\hline
\multicolumn{4}{l}{LDA present work} \\
& 0.13 & 0.70 & 1.21 \\
\hline
\multicolumn{4}{l}{GW valence pseudopotential + CPP } \\
\phantom{XX} Shirley {\it et al.}$^a$ & 1.42 & 1.75 & 1.95 \\
\hline
\multicolumn{4}{l}{GW present work} \\
 & 1.38 & 1.65 & 1.83 \\
\hline 
\multicolumn{4}{l}{GW all-electron } \\
\phantom{XX} Kotani and van Schilfgaarde$^b$ & 1.20 & 1.40 & 1.46 \\
\hline
Experiment$^c$ & 1.52 & 1.815 & 1.98 \\
\hline
\end{tabular}
\label{t.gaas}
\vspace*{2ex}

\begin{flushleft}
$^a$ Reference 9.

$^b$ Reference 6 after inclusion of spin-orbit effects.

$^c$ References 23,24.
\end{flushleft}

\end{table}

In all systems studied, we applied the GPP model \cite{h&l} in two
ways: (1) using only the valence charge density in the f-sum rule, and (2)
using a total charge density from both the valence and semicore bands. Method (1) is
physically more realistic, and predicts an energy dependence of the
inverse dielectric function that is consistent with the RPA. Method (2)
implies that semicore electrons are able to screen electric fields as
efficiently as valence electrons, which is not physical. 
Nevertheless, we find agreement between the two schemes to
better than 0.1 eV in the
converged energy gap for all systems studied. This is a consequence of
cancellation of errors: in method (2), the plasma frequency is
overestimated, and therefore the contributions $\Sigma_{sx}$ and
$\Sigma_{ch}$ are overestimated in absolute value. Since they have
opposite sign, the final quasiparticle energies are weakly
affected. We also observe that method (2) shows slower convergence of
energy gaps with respect to $n_c$, which is expected since the
$\Sigma_{ch}$ term is enhanced. All results presented in this article
were obtained using the physically more appropriate method (1).


\section{Discussion}

As shown in Figures \ref{f.si}-\ref{f.gaas}, energy gaps in the GW 
approximation have a significant dependence on the number of
unoccupied bands that is slow to converge 
in many cases. This fact was already reported in calculations
using standard pseudopotential techniques.\cite{wilkins,fleszar} 
We can analyze the physics of this convergence 
by examining the GW approximation in the static limit, the so-called
COHSEX approximation.\cite{h&l,wilkins,hedin} Bands gaps can be
calculated more easily within the COHSEX approximation and show
convergence behavior similar to the full dynamic calculation.
Under this approximation, the
Coulomb-hole term has a simple form in terms of the
polarization potential, $W_{pol}({\bf r}, {\bf r}^\prime) \equiv W({\bf r}, 
{\bf r}^\prime,\omega=0) - V({\bf  r}-{\bf r}^\prime)$,

\begin{widetext}
\begin{equation}
\Sigma_{COH} ({\bf r}, {\bf r}^\prime) = {1 \over 2} \delta({\bf r}-
      {\bf r}^\prime)  W_{pol}({\bf r}, {\bf r}^\prime) 
      = {1 \over 2} \sum_{n{\bf k}}
      \varphi_{n{\bf k}} ({\bf r}) \varphi^\star_{n{\bf k}} ({\bf
      r}^\prime) W_{pol}({\bf r}, {\bf r}^\prime) \; \; ,
\label{sigma_coh}
\end{equation}
\end{widetext}
where the second equality follows from completeness of the basis of
eigenvectors. In actual calculations, this 
sum over bands $n$ is always truncated,
and the equality is violated. 
The
Coulomb-hole energy evaluated at a given electronic state is 
calculated according to the expression

\begin{widetext}
\begin{equation}
\bra m {\bf k} | \Sigma_{COH} | m {\bf k} \ket = {1 \over 2}
\sum_{\bf GG' q} W_{\bf GG'} ({\bf q}) \sum_n \left[ {\cal
  M}^{nm}_{\bf G} 
({\bf k,q}) \right]^\star {\cal M}^{nm}_{\bf G'}({\bf k,q})\; \;,
\label{k_sigma_coh}
\end{equation}
\end{widetext}
where $W_{\bf GG'} ({\bf q})$ are the coefficients in the plane-wave 
expansion of the polarization potential, \cite{h&l,wilkins} and
we define

\begin{equation}
{\cal M}^{nm}_{{\bf G}}({\bf k,q}) = \bra n {\bf k - q} | {\rm e}^{-i 
{\bf (q + G) \cdot r } } | m {\bf k} \ket \; \; .
\label{mat_el}
\end{equation}

Physically, the summation in
Eq. (\ref{sigma_coh}) describes virtual transitions
produced when the quasiparticle induces a charge fluctuation
around itself. $\Sigma_{COH}$ is the energy associated to the
interaction between the quasiparticle and the induced
charge fluctuation. The matrix elements, Eq. (\ref{mat_el}), decay
slowly as the energy difference between bands $m$ and $n$ increases.
Monitoring the convergence of these matrix elements
provides a good estimate of the relative error in the $\Sigma_{COH}$, but
gauging the absolute convergence of the self-energy requires knowing
$W_{pol}$, which depends on the physical system.

Although we have not directly investigated the importance of imposing
self-consistency in the calculation of the self-energy, this is an
unsetled issue and we address it briefly below.
Von Barth and Holm have
investigated the effect of self-consistency in the electron gas,\cite{holm}
and concluded that restricted self-consistency has
small but significant effect on the full bandwidth and in the satellite
structure of the electron gas. On the other hand, full self-consistency gives
a poor description of the satellite structure and the bandwidth is
drastically increased. Inclusion of vertex corrections 
are expected to recover the good, non-self-consistent results, but
calculation of vertex corrections is not a simple task even for 
the electron gas
system.\cite{wilkins,holm} Self-consistency has been recently applied to real
materials,\cite{ku,eguiluz} and the valence bandwidth is also shown to 
increase when self-consistency is imposed. 
It appears that one must 
therefore include self-consistency and vertex corrections together in
order to obtain a meaningful picture.
Inclusion of vertex corrections and self-consistency is tangential
 for the purpose of our work, which is to compare 
pseudopotential-based and all-electron GW calculations and understand 
the role of semicore electrons.


\section{Conclusion}

We conclude that for the systems considered, the valence-only
pseudopotential method does not suffer from large errors from the neglect
of core states, as claimed in some all-electron
calculations. \cite{kotani,ku} While semicore effects are negligible in Si, 
they are important if one aims at good quantitative agreement with 
experiment in
Ge and GaAs.\cite{wilkins,aryasetiawan_rev,aryasetiawan_prb,rohlfing} 
However, discrepancies between pseudopotential and all-electron based GW
calculations reported in recent works \cite{kotani,ku} 
may be explained
by a lack of numerical convergence in the latter.
Specifically, the self-energy calculated within the GW method
has slow convergence with respect to the number of energy bands
included in the calculation of the Green's function, as 
is demonstrated in this work and has been pointed out in
the past. \cite{wilkins,fleszar} This convergence behavior is
present in the static limit to GW, the COHSEX approximation, and can
be analyzed by comparing the COHSEX Coulomb-hole energy, Eq. (\ref{sigma_coh}),
obtained with and without explicit summation over energy bands.

\begin{acknowledgments}
We thank Andrew Canning and David Clatterbuck for fruitful
discussions regarding the FLAPW method. This work was supported by
National Science Foundation Grant No. 
DMR00-87088 and by the Director, Office of Science, Office of Basic Energy
Sciences, Division of Materials Sciences and Engineering, U.S. Department
of Energy under Contract No. DE-AC03-76SF00098.  Computational resources
have been provided by NSF at the National Partnership for Advanced
Computational Infrastructure, and DOE at the National Energy Research
Scientific Computing Center. 
\end{acknowledgments}




\begin{references}

\bibitem{h&l} M.S. Hybertsen and S.G. Louie, Phys. Rev. B {\bf
34}, 5390 (1986).

\bibitem{wilkins} W.G. Aulbur, L. J\"onsson, and J.W. Wilkins, {\it
  Solid State Physics}, eds. F. Seitz, D. Turnbull, and H. Ehrenreich,
  Academic, New York,  vol. {\bf 54}, 1 (2000). 

\bibitem{aryasetiawan_rev} F. Aryasetiawan and O. Gunnarsson,
  Rep. Prog. Phys. {\bf 61}, 237 (1998).

\bibitem{blochl} P.E. Bl\"ochl, Phys. Rev B {\bf 50}, 17953 (1994).

\bibitem{hamada} N. Hamada, M. Hwang, and A.J. Freeman,
  Phys. Rev. B {\bf 41}, 3620~(1990).

\bibitem{kotani} T. Kotani and M. van Schilfgaarde,
Sol. State. Comm. {\bf 121}, 461 (2002).

\bibitem{ku} W. Ku and A.G. Eguiluz,  Phys. Rev. Lett. {\bf 89}, 126401
(2002).

\bibitem{lebegue} S. Leb\`egue, B. Arnaud, M. Alouani, and
  P.E. Bl\"ochl, Phys. Rev B {\bf 67}, 155208 (2003).

\bibitem{szl}  E.L. Shirley, X. Zhu, and S.G. Louie,
  Phys. Rev. B {\bf 56}, 6648 (1997).

\bibitem{ihm} R.M. Dreizler and E.K.U Gross, {\it Density functional
  theory: an approach to the quantum many-body problem}
  (Springer-Verlag, Berlin, 1990).

\bibitem{p-z} J.P. Perdew and A. Zunger, Phys. Rev. B {\bf 23}, 5048 (1981).

\bibitem{tm2} N. Troullier and J.L. Martins, Phys. Rev. B {\bf
43}, 1993 (1990).

\bibitem{kerker} G.P. Kerker, J. Phys. C {\bf 13}, L189 (1980).

\bibitem{rohlfing} M. Rohlfing, P. Kr\"uger, and J. Pollmann,
  Phys. Rev. B {\bf 57}, 6485 (1998).

\bibitem{hedin} L. Hedin and S. Lundqvist, {\it Solid State Physics},
eds. F. Seitz, D. Turnbull, and H. Ehrenreich,
  vol. {\bf 23}, 1, Academic, New York (1969).

\bibitem{k-b} L. Kleinman and D. M. Bylander, Phys. Rev. Lett. {\bf
  48}, 1425 (1982).

\bibitem{m-pack} H.J. Monkhorst and J.D. Pack, Phys. Rev B {\bf 13},
  5188 (1976).

\bibitem{h&l-so} M.S. Hybertsen and S.G. Louie, Phys. Rev. B {\bf 34},
  2920 (1986).

\bibitem{aryasetiawan_prb} F. Aryasetiawan and O. Gunnarsson,
  Phys. Rev. B {\bf  54}, 17564 (1996).

\bibitem{fleszar} A. Fleszar and W. Hanke, Phys. Rev. {\bf 56}, 10228
  (1997).

\bibitem{holm} U. von Barth and B. Holm, Phys. Rev. B {\bf 54}, 8411
  (1996); Phys. Rev. B {\bf 55}, 10120 (1997); B. Holm and U. von
  Barth, Phys. Rev. B {\bf 57}, 2108 (1998).

\bibitem{eguiluz} W-D. Sch\"one and A.G. Eguiluz,
  Phys. Rev. Lett.{\bf 81}, 1662 (1998).

\bibitem{l-b87} Landolt-B\"ornstein, {\it Numerical Data and
  Functional Relationships in Science and Technology},
  eds. K.-H. Hellwege, O. Madelung, M. Schulz, and H. Weiss, New
  Series, vol III, Pt. 22a, Springer-Verlag, New York (1987).

\bibitem{laut} P. Lautenschlager, M. Garriga, S. Logothetidis, and
  M. Cardona, Phys.  Rev. B {\bf 35}, 9174 (1987).

\end{references}
\end{document}